\documentclass[12pt]{article}
\usepackage[margin=1in]{geometry}
\usepackage{graphicx} % Required for inserting images
\usepackage{aas_macros}
\usepackage[numbers]{natbib}
\usepackage{amsmath,amssymb}
\usepackage{xspace}

\def\roman{\textit{Roman}\xspace}
\def\gaia{\textit{Gaia}\xspace}

\def\msun{${\rm{M}_\odot}$}
\newcommand{\e}[1]{\times 10^{#1}} 

\usepackage[compact]{titlesec}

\usepackage{hyperref}
\hypersetup{colorlinks,bookmarksnumbered,
linkcolor=cyan,citecolor=cyan,urlcolor=cyan}
\usepackage[usenames, dvipsnames]{xcolor}

\begin{document}

% \maketitle

\noindent {\Large{Roman CCS White Paper}}
\section*{Gravitational Wave Detection with Relative Astrometry using Roman's Galactic Bulge Time Domain Survey}
\textbf{Roman Core Community Survey}: Galactic Bulge Time Domain Survey\\

\noindent \textbf{Scientific Categories}: supermassive black holes and active galaxies\\

\noindent \textbf{Additional scientific keywords}: gravitational waves, supermassive black holes, quasars\\

\noindent \textbf{Submitting Author}\\
\indent Name: Kris Pardo \\
\indent Affiliation: University of Southern California \\
\indent Email: \href{mailto:kmpardo@usc.edu}{kmpardo@usc.edu}\\

\noindent \textbf{List of contributing authors (including affiliation and email)}:\\ Tzu-Ching Chang (JPL/Caltech; \href{mailto:tzu-ching.chang@jpl.nasa.gov}{tzu-ching.chang@jpl.nasa.gov}),\\ Olivier Dor\'{e} (JPL/Caltech; \href{mailto:olivier.p.dore@jpl.nasa.gov}{olivier.p.dore@jpl.nasa.gov}) \\
Yijun (Ali) Wang (Caltech; \href{mailto:yijunw@caltech.edu}{yijunw@caltech.edu})\\

\noindent \textbf{Abstract}:
Gravitational waves (GWs) are a new avenue of observing our Universe. So far, we have seen them in the ~10-100 Hz range, and there are hints that we might soon detect them in the nanohertz regime. Multiple efforts are underway to access GWs across the frequency spectrum; however, parts of the frequency space are currently not covered by any planned or future observatories. Photometric surveys can bridge the microhertz gap in the spectrum between LISA and Pulsar Timing Arrays (PTAs) through relative astrometric measurements. Similar to PTA measurements, these astrometric measurements rely on the correlated spacetime distortions produced by gravitational waves at Earth, which induce coherent, apparent stellar position changes on the sky. To detect the
 microhertz GWs with an imaging survey, a combination of high relative astrometric precision, a large number of observed stars, and a high cadence of exposures are needed. Roman's proposed core community survey, the Galactic Bulge Time Domain Survey (RGBTDS), would have all of these components. RGBTDS would be sensitive to GWs with frequencies ranging from $7.7\e{-8}$ Hz to  $5.6\e{-4}$ Hz, which opens up a unique GW observing window for supermassive black hole binaries and their waveform evolution. We note that small changes to the survey could enhance Roman's sensitivity to GWs, making it possible to observe the GW background signal that PTAs have recently hinted at with an SNR $\sim$ 70.

\newpage

\subsection*{Introduction}
\vspace{-0.1cm}
\noindent The detection of GWs is one of the major scientific breakthroughs of the last decade, and the science we have done with them so far is only a small glimpse of their full potential. GWs have allowed us to probe populations of stellar mass black holes, neutron star interiors, test theories of gravity, and measure the Hubble constant \citep{GW150914, GW170817, GW170817_NSEOS, GW170817GRTests, Baker2017, Ezquiaga2017,  GW170817_H0}. So far, we have only definitively observed GWs in the ${\sim} 10-1000~\rm{Hz}$ band. Like electromagnetic waves, GWs can be produced at many different frequencies, which allow us to probe a variety of physics. The millihertz band, which will be observed by the Laser Interferometer Space Antenna (LISA), allows us to probe $10^3-10^7~$\msun\ (where \msun\ denotes a solar mass) black hole mergers \citep{LISA}. \roman's ability to find precursors to LISA binaries and provide other complementary science is discussed in a separate white paper \citep{Haiman2023}. The nanohertz band, currently monitored by pulsar timing arrays (PTAs) \citep[e.g.,][]{NANOGRAV, EPTA}, probes the gravitational wave background (GWB) produced by the superposition of all inspiraling supermassive black holes in the Universe \citep{Phinney2001}. Excitingly, several PTA collaborations have reported a possible observation of this GWB \citep{NGGWBhint, EPTAGWBhint, PPTAGWBhint, IPTAGWBhint}. 

% They do not yet have statistically significant evidence that the expected spatial correlations are there, and thus they cannot yet claim a detection. However, it is expected that they will make a detection in the next ${\sim}$2 years, if the signal is indeed at the amplitude they have already seen \citep{Mingarelli2019}.

However, there are gaps in the gravitational wave spectrum where there are no planned experiments. One of the most conspicuous gaps is the microhertz band, between LISA and PTAs. This band is where the largest supermassive black hole (SMBH, $10^8-10^{10}$~\msun) binaries, which create the GWB that PTAs see, actually coalesce, and where we could see the GWB produced by the inspiraling of slightly lighter black hole binaries ($10^6-10^8~$\msun). Observing this band would also give us insight into the populations that LISA would later see merge. To date, the only constraints in this frequency regime come from the \textit{Cassini} spacecraft, which found an upper limit on the GWB at $f=0.3~\rm{mHz}$ of $h_c < 2\times 10^{-15}$ \citep{Armstrong2003}. There are some proposals for how to use other objects and methods to observe these GWs \citep{Fedderke2021, Bustamante-Rosell2022, Sesana2021}; however, these will either require significant investment into new missions or will require waiting until after LISA launches (currently slated for 2037 \citep{LISA_webpage}).

\subsection*{Detecting Gravitational Waves with Photometric Surveys} 
\begin{figure}[ht]
    \begin{center}
    % \begin{boxedminipage}{\textwidth}  
    \centering
          \includegraphics[width=0.85\textwidth]{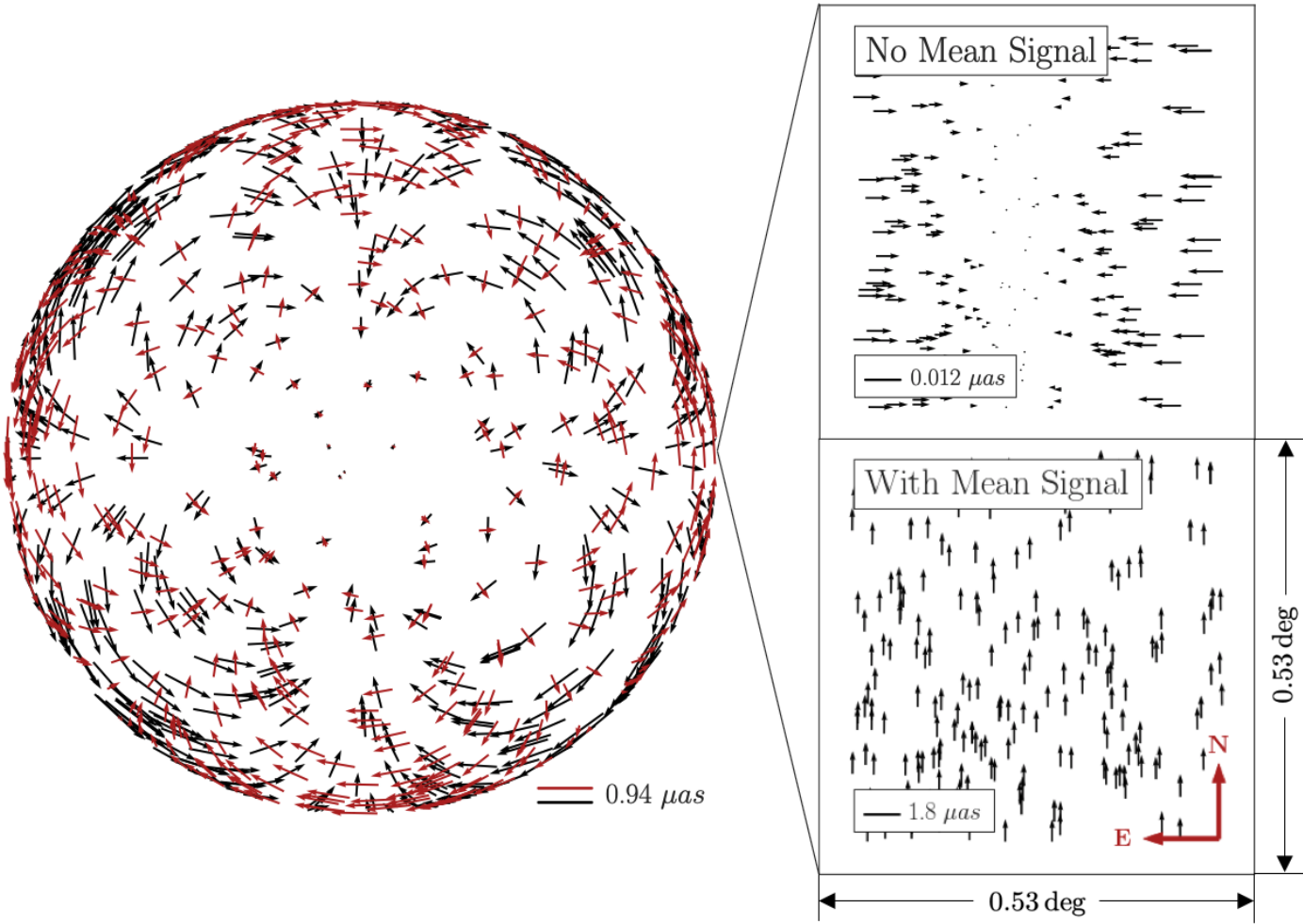}
      \caption{Illustration of the expected stellar astrometric deflections. Left: Orthographically projected GW deflections from a $10^9$ \msun equal-mass SMBH binary at 1 Mpc at the center. The arrows denote the maximal change in apparent position produced by a GW with frequency ${\sim} 10^{-6}~\rm{Hz}$. Black arrows correspond to the real part of the waveform at GW phase $\phi$ = 0 (plus polarization), and the red arrows correspond to that at $\phi = \pi / 4$ (cross polarization). Right: Simulated deflections within \roman's FOV between two expsorues of the RGBTDS. The lower panel shows the total deflection, and the upper panel shows the deflection after subtracting the mean, since the mean is expected to be absorbed in the pointing reconstruction. Reproduced from Ref.~\cite{Wang2021}.}
    \label{fig:deflections}
    \vspace{-0.5cm}
    % \end{boxedminipage}
    \end{center}
\end{figure}

\vspace{-0.1cm}
Astrometry is one possible way to observe GWs and fill the gaps in the spectrum \citep{Book2011, Moore2017, Klioner2018}. Similar to PTAs, this method works by looking for correlated distortions produced by a GW passing through the Earth and distorting the spacetime at Earth. PTAs look for the time delay or early arrival of radio signals, which occur when GWs distort distances along the line of sight to pulsars. The astrometric measurements rely on the correlated spacetime distortions produced by gravitational waves at Earth, which induce coherent, apparent stellar position changes on the sky (Figure~\ref{fig:deflections}). Many papers have considered looking for astrometric GWs with \gaia \citep{Moore2017, Klioner2018}; however, \gaia does not have the frequency resolution (\textit{i.e.}, observing cadence) to cover the microhertz band. In addition, \gaia's data processing, which subtracts out large-scale correlated motions on the sky, would make searching for GWs particularly challenging. However, recent work \citep[][]{Wang2021,Wang2022} has proposed instead using relative astrometry to search for GWs -- this allows \textit{any} photometric survey to be used as a GW probe, provided it has sufficient relative astrometric resolution, a large number of surveyed stars, and frequent exposures of the same region of the sky. {\bf The Roman Galactic Bulge Time Domain Survey (RGBTDS) \citep{Gaudi2019} would have all of the components necessary to make \roman a GW probe.}
In the following, we briefly explore the survey parameters that would enable this science.

\subsection*{Survey Design Considerations}
The amplitude of a GW, typically called the ``strain'', $h$, is a measure of the fractional change in length caused by a GW. However, for typical sources (e.g., the stochastic GWB), the relevant strain for estimating the sensitivity is typically the characteristic strain, $h_c$, which can be related to the power spectrum of a given $h(t)$ signal. For an astrometric survey, we can approximate the characteristic strain sensitivity as \citep{Wang2022}:
\begin{equation}
    h_c \sim \sigma_\theta \sqrt{\frac{2fT_{\rm{cad}}}{N_s N_{\rm{exp}}}} \; ,
\end{equation}
where $\sigma_\theta$ is the dimensionless relative astrometric precision, $N_s$ is the number of stars observed, $N_{\rm{exp}}$ is the number of exposures, $f$ is the frequency, and $T_{\rm{cad}}$ is the time between exposures.

Note that the sensitivity relies heavily on the number of stars and the number of exposures. Because photometric surveys often have flexible pointing directions and exposure times, this makes them easily able to increase their GW sensitivity. In addition, searches for transiting exoplanets often try to maximize these two survey characteristics. To target certain regions in frequency space in the GW spectrum, a survey needs to have an appropriate exposure frequency. For a survey with a cadence of $T_{\rm{cad}}$ and total survey duration $T_{\rm{survey}}$, the corresponding GW frequencies are:
\begin{equation}
    \frac{1}{T_{\rm{survey}}} \lesssim f \lesssim \frac{1}{2T_{\rm{cad}}}.
\end{equation}

Because the frequency is set by the relatively flexible cadence and survey times, this makes \roman a versatile probe of GWs. For generic surveys with $T_{\rm{cad}} \sim 10~\rm{min} - 1~\rm{hour}$ and several year baselines, this provides sensitivity to the GW spectrum between LISA and PTA, which is currently inaccessible through any direct GW experiments (see Figure~\ref{fig:hc}).

\subsection*{The ``mean'' deflection signal}

We note that \roman, like most telescopes, takes relative astrometric measurements, recording only the relative positions of objects with each other and across the exposures with its nominal astrometric resolution. The absolute positions will be determined by referencing guide stars, and by the telescope pointing with larger uncertainty. Consequently, the pointing reconstruction strategy of \roman will likely absorb deflections uniform across the field of view (FoV). We refer to this almost uniform deflection as the FoV mean signal, which is by unobservable for FoV smaller than the correlation length of the on-sky GW signal.

The right panels of Figure~\ref{fig:deflections} show the astrometric deflection in an approximate \roman FoV, assuming the telescope is in the Galactic plane and points directly to the Galactic center. This FoV model has roughly the same area as the true FoV of \roman, but differs in shape. We adopt it here for simplicity, and note that it does not significantly change our results here.
The bottom panel shows the total deflection pattern while the upper panel shows the deflection pattern after subtracting the mean deflection. If there are no changes to the RGBTDS, this is expected to be the actual observed signal, which is about 100 times smaller in strain amplitude compared to the mean signal \cite{Wang2021}.

There are a few mitigation strategies that could be employed here. The first would be to try to subtract the mean signal using onboard instrument readings. This is worth considering closely. Another strategy would involve slightly modifying the RGBTDS. Currently, the RGBTDS involves monitoring 7 fields, with exposures of each field interleaved. This means that every single exposure would be subject to the mean subtraction issue we mention here. \textbf{If instead, only one field is used and Roman stared at the one field contiguously over each 72 day period, this would cut down on the mean subtraction issue considerably.} There would still be some mean subtraction, since \roman would likely need to stabilize its pointing after drifting for some amount of time. However, this would enable full sensitivity to GW signals at frequencies higher than this stabilization frequency. Ignoring any new stars that could be observed by going deeper in this one field, there would be some trade-off given that fewer stars would be observed. However this is an $\mathcal{O}(1)$ change in sensitivity compared to the $\mathcal{O}(100)$ increase in sensitivity from keeping the mean signal.

\subsection*{Prospects for detecting GWs with \roman}
\label{subsec:snr}

\noindent The currently planned RGBTDS remarkably has each of the three key components needed for a successful astrometric detection of GWs:
\begin{itemize}
    \item Large Number of Stars Surveyed: a total of N$_s \sim 10^8$ stars with W145$_{AB} < 23$ \cite{Gaudi2019}.
    \item Frequent and Numerous Exposures of the Same Region: a 15-minute observing cadence with six 72-day observational seasons spread out over the nominal 5-year mission time is expected. These constraints then give a conservative frequency range as:
$7.7 \times 10^{-8}$ Hz $< \Omega < 5.6 \times 10^{-4}$ Hz, a unique window in the GW spectrum. Additionally, each of these $\sim 10^8$ stars would be imaged $\sim 41,000$ times, yielding high sensitivity to GWs.

    \item High Relative Astrometric Precision: \roman's relative astrometry has a precision of 1.1~\rm{mas} for a single exposure, estimated for H$_{AB} = 21.6$ stars \cite{WFIRST2019}.
\end{itemize}

\begin{figure}[!t]
\begin{center}
    \includegraphics[width=0.9\textwidth]{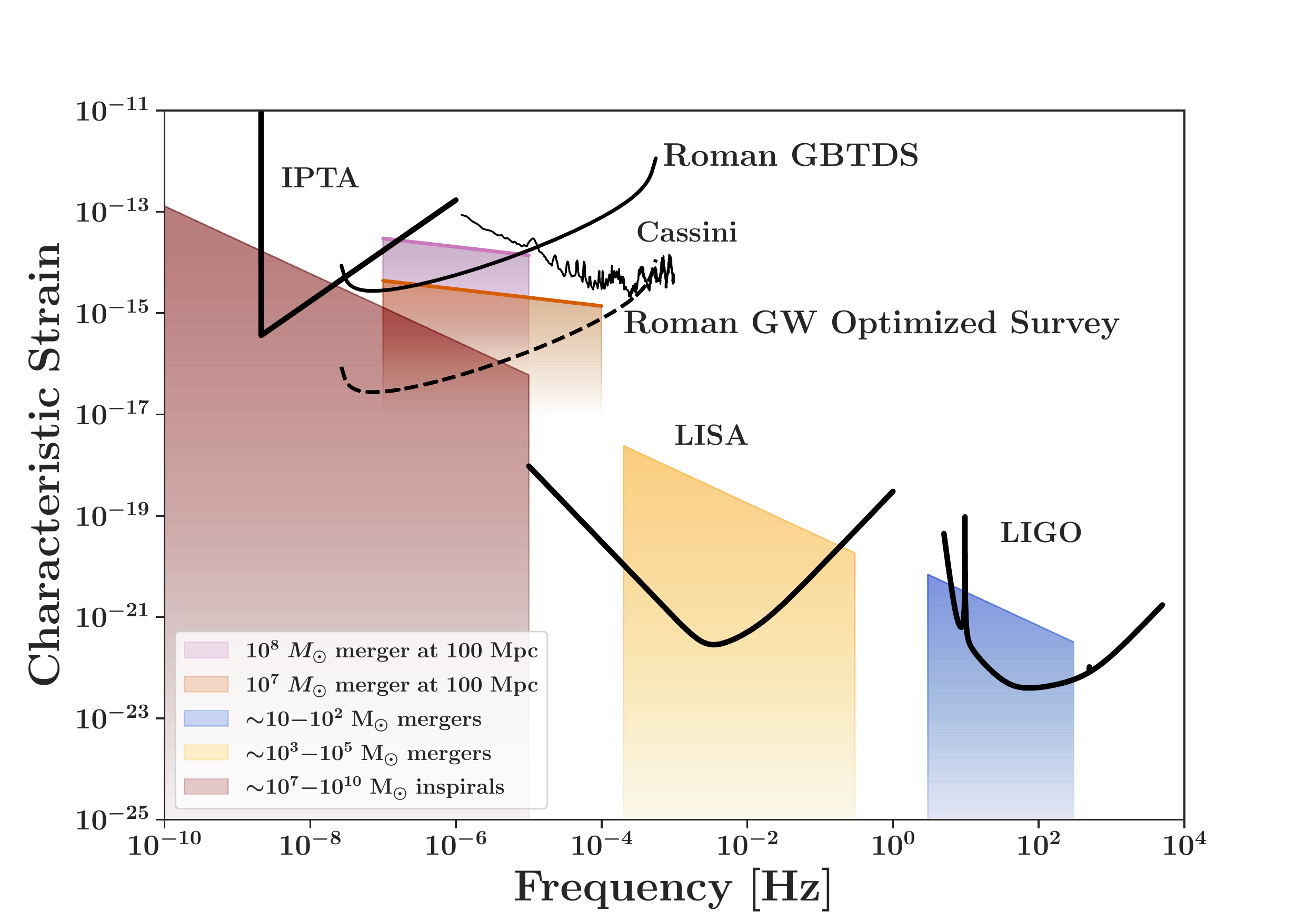}
  \caption{Projected characteristic strain, $h_c$, sensitivity for Roman RGBTDS, the GW optimized survey we describe here, and other experiments and signals.
  }
  \label{fig:hc}
  \vspace{-0.7cm}
\end{center}
\end{figure}

\noindent These features give \roman a frequency sensitivity perfectly between LISA and PTA, and a high sensitivity to GW strain (see Fig.~\ref{fig:hc}). However, previous analytic estimates \citep{Wang2021, Wang2022} suggest that the GW strain amplitudes would be considerably curbed if the mean signal could not be recovered, as discussed above.

If the optimal survey discussed above occurs, which could recover the mean signal, then \roman would be capable of observing single supermassive black hole binaries with mass $\mathcal{M}_c \sim 10^9$~\msun\ at signal-to-noise ratio SNR$\geq 3$ out to a luminosity distance $d_L \sim 1~\rm{Gpc}$. In addition, \textbf{\roman would be able to detect the stochastic GWB with an estimated $\mathrm{SNR}\sim 70$, if the expected signal is at the level currently hinted at by various PTA collaborations, $h_c(f) = 2.8\e{-15}(f/1~\rm{yr})^{-2/3}$ \citep{NGGWBhint, EPTAGWBhint, PPTAGWBhint, IPTAGWBhint}.} If we cannot recover the mean, \roman would still be sensitive to $\mathcal{M}_c \sim 10^9$~\msun\ at signal-to-noise ratio SNR$\geq 3$ out to a luminosity distance $d_L \sim 10~\rm{Mpc}$, which is not a large enough volume to expect sources this massive. However, \roman would still be sensitive to the stochastic GWB with an estimated $\mathrm{SNR}\sim 1$ and set interesting and unique limits on SMBH populations and evolution.

\newpage

\bibliography{ref}

\end{document}